\documentstyle[prd,preprint,aps,eqsecnum]{revtex}

\begin{document}
\draft
\title{On quark mass correction to the string potential
}

\author{G. Germ\'an}
\address{Instituto de F\'{\i}sica, Laboratorio de Cuernavaca,\\
Universidad Nacional Aut\'onoma de M\'exico,\\
Apartado Postal 48-3, 62251 Cuernavaca, Morelos, M\'exico\\}

\author{Yu\ Jiang}
\address{Facultad de Ciencias,
Universidad Aut\'onoma del Estado de Morelos,\\
Av. Universidad 1001, Col. Chamilpa,\\
62210 Cuernavaca, Morelos, M\'exico}

\date{April 1997}     
\maketitle
\begin{abstract}
Following recent work by Lambiase and Nesterenko we study in detail the
interquark potential for a Nambu-Goto string with point masses attached 
to its ends. We obtain exact solutions to the gap equations for the Lagrange
multipliers and metric components and determine the potential without 
simplifying assumptions. We also discuss L\"{u}scher term and argue that it 
remains universal.
\end{abstract}

\section{Introduction}

There has been considerable effort in trying to understand the forces between
quarks in terms of strings and several models have been proposed with different
degrees of success. The Nambu-Goto model [1], which is a direct generalization 
of the covariant action for a relativistic point particle moving in space-time,
describes the evolution of a string. When this string evolves sweeps out a 
two-dimensional world sheet surface embedded in a higher dimensional space-time.
The area of this surface is precisely the Nambu-Goto action. Close to this are the 
generalized Eguchi models [2] of which Schild's [3] is a particular case. The
functional-integral quantization of these models has been studied by L\"{u}scher,
Symanzik and Weisz [4] and by Alvarez [5] who calculated the static potential in
the large-d limit, where d is the number of dimensions of the embedding space.
The result obtained by Alvarez turned out to be correct for any d as shown latter by
Arviz [6]. The Nambu-Goto string model gives qualitatively encouraging results as a 
large-N QCD string i.e., the interquark potential is linear for large distances which 
is understood as a signal of confinement, it also has linear Regge trajectories and 
presents a transition to a deconfined phase with vanishing string tension at 
certain temperature [7]. Quantitatively, however, this model is not in very good
shape with numerical values closer to those obtained in Monte Carlo simulations
of an SU(2) lattice gauge theory rather than SU(3) [8]. Also it has been shown [9] 
that agreement of the Nambu-Goto string with a calculation of the high-temperature
partition function of a QCD flux tube would require an infinite number of massive 
world-sheet degrees of freedom. Thus, the model has been modified by populating 
the string with scalar an Fermi fields improving some of the quantitative 
results [10]. However the prize to pay is too high since the conformal invariance
of the theory is explicitly broken [11]. It seems that the Nambu-Goto string or 
naive modifications of it will not give us the QCD string. Still the Nambu-Goto 
model remains very useful as the simplest string model where some calculations
can be done without undue effort and mathematical methods as well as new physical
ideas can be tested. More elaborated extensions of Nambu-Goto have incorporated 
an extrinsic curvature term in the action (the so-called rigid or Polyakov-Kleinert
string [12]) and several properties of interest have been investigated [13].
Rigid strings coupled to long range Kalb-Ramond fields have been also studied 
[14], and more recently "confining strings" [15] seem to be very promising
models for the QCD string. In all of these models one important problem is to 
determine the potential between two sources i.e., the so called interquark static 
potential. This potential has been calculated by various perturbative and
non-perturbative methods. The common feature has been, however, the assumption 
of infinitely massive quarks at the ends of the string which is equivalent to
impossing fixed ends boundary conditions. In a recent series of papers a 
consistent method has been proposed to study the effects of finite point masses
attached to the ends of the string [16]. In particular a variational estimation 
of the Nambu-Goto string potential has been worked out although with some 
simplifying assumptions[17]. Here we reconsider this problem and present the 
solutions to the gap equations and determine the interquark potential as 
well as other quantities of interest exactly. We also provide a discussion of 
L\"{u}scher term and argue that it remains universal with no mass contributions
coming from the point particles attached to the ends of the string. In section II
we present the model and equations for the Lagrange multipliers and metric 
components. We also obtain a very simple looking formula for the interquark 
potential. The numerical analysis of the equations and various quantities of 
interest is carried out in section III. We also compare with the approximated 
results of [17]. Finally section IV comprises a discussion of L\"{u}scher term and 
argue that it remains universal. We conclude with a brief account of our results.

\section{The model and gap equations}

At the quantum level the Nambu-Goto model is given by the following functional 
integral
\begin{equation}
Z=\int [Dx^{\mu}] e^{-S},
\end{equation}
in Euclidean space, the action S is
\begin{equation}
S=M_0^2\int d^{2}\xi\sqrt{g}+\sum_{a=1}^2m_a\int_{C_i}ds_a,
\end{equation}
where $M_0^2$ is the string tension, $C_i (i=1,2)$ are the world trajectories 
of the string massive ends and $g$ is the determinant of the metric
\begin{equation}
g_{ij}=\partial_{i} x^{\mu}(\xi_{i})\partial_{j} x^{\nu}(\xi_{i})\eta _{\mu\nu}, 
\qquad i=0,1.
\end{equation}
The $x^{\mu}$, $\mu=0,1,...,d-1$ are the string coordinates and $\eta_{\mu\nu}$
is the embedding Euclidean metric of the space where the string evolves, 
$g^{ij}$ is thus the induced metric on the world sheet swept out by the string. 
To study the model further it is convenient to specify a gauge, we choose the 
"physical gauge" or Monge parametrization
\begin{equation}
x^{\mu}(\xi_{i})=(t,r,u^{a}(t,r)),
\end{equation}
where the $\vec{u}^a(t,r)$, $a=2,...,d-1$ are the $(d-2)$ transverse oscillations of the 
string. We further introduce composite fields $\sigma_{ij}$ given by
\begin{equation}
\sigma_{ij}=\partial_{i} \vec{u}\cdot\partial_{j} \vec{u}.
\end{equation}
The metric $g_{ij}$ and string coordinates $\vec{u}$ become independent fields by 
introducing Eq.~(2.3) as a constraint. This requires the use of Lagrange 
multipliers $\alpha^{ij}$ which also become independent variables.
The functional integral Eq.~(2.1) then becomes
\begin{equation}
Z = \int [D\vec{u}][D\alpha][D\sigma] e^{-S(\vec{u},\alpha,\sigma)},
\end{equation}
where the action Eq.~(2.2) is now given by
\begin{eqnarray}
S &=& M_0^2\int_0^\beta dt\int_0 ^R dr [\sqrt{det(\delta_{ij} + \sigma_{ij})} + 
\frac{1}{2}\alpha^{ij}(\partial_{i} \vec{u}\cdot\partial_{j} \vec{u} - 
\sigma_{ij})]
\nonumber \\
 &&+ \sum_{a=1}^2 m_a\int dt\sqrt{1+\dot{\vec{u}^2}(t,r_a)} \qquad r_1=0, \qquad r_2=R.
\end{eqnarray} 
It has been shown by Alvarez that, at the saddle point, the Lagrange parameters 
$\alpha^{ij}$ as well as the metric components $\sigma_{ij}$ become symmetric constant 
matrices with no dependence on $t$ and $r$. Thus while $\vec{u} = \vec{u}(t,r)$ is in general a 
function of $t$ and $r$, $\dot{\vec{u}^2} = \sigma_0$ becomes, at the saddle point, a constant. This
fact simplifies the problem considerably. Since the action is quadratic in the string
oscillations $\vec{u}^a$ we can do the gaussian integral inmediately. The resulting action,
in the particular case where $m_1 = m_2 = m$, can be written as
\begin{equation}
S(\alpha,\sigma) = M_0^2\beta R[\sqrt{(1+\sigma_0)(1+\sigma_1)} - \frac{1}{2}(\alpha_0\sigma_0+
\alpha_1\sigma_1)-\sqrt{\frac{\alpha_1}{\alpha_0}}\lambda]+2m\beta.
\end{equation}
Here $\lambda$ is related to the Casimir energy $E_c = \frac{1}{2} \sum_{k=1}^{\infty}\omega_k$ as follows
\begin{equation}
\lambda=-\frac{(D-2)}{M_0^2R}E_c,
\end{equation}
and the last term in Eq.~(2.8) is the contribution to the action due to the point 
masses at the ends of the string. This term can be set to zero with an appropriate redefinition 
of $S$. Thus we ignore this term in what follows. The Casimir energy $E_c$
depends on the eigenmomenta $\omega_k$ which on its turn depend on the boundary 
conditions imposed on the system. For a string with infinitely heavy quarks 
attached to its ends we impose fixed ends boundary conditions, in this case
\begin{equation}
\omega_k=\frac{n\pi}{R} \qquad n=1,2,...,
\end{equation}
and the Casimir energy is
\begin{equation}
E_c=\frac{1}{2}\sum_{k=1}^{\infty} \omega_k=\frac{\pi}{2R}\sum_{n=1}^{\infty} n=-\frac{\pi}{24R},
\end{equation}
where the last term was obtained by the use of Riemann's $\zeta$-function i.e.,
$\sum_{n=1}^{\infty} n=[\sum_{n=1}^{\infty} \frac{1}{n^{\nu}}]_{\nu=-1}=\zeta (-1)=-\frac{1}{12}.$
In the case of finite quark masses the problem becomes increasingly difficult to
dealt with even when $m_1 = m_2$. It can be shown that in this case $(m_1=m_2=m)$
the Casimir energy is given by [17]
\begin{equation}
E_c=\frac{1}{2\pi R}\int_0^{\infty} dx\ln[1-(\frac{x-s}{x+s})^2 e^{-2x}],
\end{equation}
where
\begin{equation}
s=\frac{\rho}{\mu}\alpha_0\sqrt{1+\sigma_0},
\end{equation}
and
\begin{equation}
\rho=M_0R, \qquad \mu=\frac{m}{M_0},
\end{equation}
are dimensionless quantities corresponding to the (extrinsic) 
length and point masses attached to the ends of the string,
respectively. The equation for $\lambda$ Eq.~(2.9) becomes
\begin{equation}
\lambda = -\frac{(D-2)}{2\pi\rho^2}\eta (s),
\end{equation}
where
\begin{equation}
\eta (s)=\int_0^{\infty} dx\ln[1-(\frac{x-s}{x+s})^2 e^{-2x}].
\end{equation}
It is also convenient to write $\lambda$ in the form
\begin{equation}
\lambda =\frac{(D-2)\pi}{24\rho^2}-\frac{(D-2)}{2\pi \rho^2}\int_0^{\infty} dx\ln[1+
\frac{4sx}{(x+s)^2}\frac{1}{e^{2x}-1}].
\end{equation}
Note that $\lambda$ is a function of $\alpha_0$ and $\sigma_0$ through $s$,
Eq.~(2.13). Thus when writing the equations for the Lagrange multipliers and metric
components derivatives of $\lambda$ with respect to $\sigma_0$ and $\alpha_0$
should appear. These are given by
\begin{mathletters}\label{2.18}
\begin{eqnarray}
\alpha_0&=&\sqrt{\frac{1+\sigma_1}{1+\sigma_0}}-\frac{\sqrt{\alpha_0\alpha_1}}{1+\sigma_0}
\frac{\partial\lambda}{\partial\alpha_0},\\
\alpha_1&=&\sqrt{\frac{1+\sigma_0}{1+\sigma_1}},\\
\sigma_0&=&\frac{1}{\alpha_0}\sqrt{\frac{\alpha_1}{\alpha_0}}\lambda-2\sqrt{\frac{\alpha_1}{\alpha_0}}
\frac{\partial\lambda}{\partial\alpha_0},\\
\sigma_1&=&-\frac{1}{\sqrt{\alpha_0\alpha_1}}\lambda,
\end{eqnarray}
\end{mathletters}
where, in Eq.~(2.18a), $\frac{\partial\lambda}{\partial\sigma_0}$ has been replaced by
\begin{equation}
\frac{\partial\lambda}{\partial\sigma_0}=
\frac{\alpha_0}{2(1+\sigma_0)}\frac{\partial\lambda}{\partial\alpha_0}.
\end{equation}
The potential $V(\rho)$ is obtained in the usual
way $e^{-{\beta V(\rho)}}\sim Z$, $\beta \rightarrow \infty$ and is given by 
the simple looking formula
\begin{equation}
\overline{V}(\rho)=\rho\alpha_0,
\end{equation}
which follows from Eq.~(2.8) and the gap equations (2.18). The potential $\overline{V}(\rho)$
is also a dimensionless quantity, $\overline{V}(\rho) = M_0^{-1}V(\rho)$. Of course there is no 
way to solve Eqs.~(2.18) analytically thus Eq.~(2.20) is only a formal expression
for $\overline{V}(\rho)$. One can play with Eqs.~(2.18) and write down an expression for
$\alpha_0$
\begin{equation}
\alpha_0=\sqrt{1-\frac{1+\alpha_0\alpha_1}{\sqrt{\alpha_0\alpha_1}}\lambda-
(1-\alpha_0\alpha_1)\sqrt{\frac{\alpha_0}{\alpha_1}}\frac{\partial\lambda}{\partial\alpha_0}},
\end{equation}
which will be useful for discussing some limiting situations in the last section.

\section{Numerical analysis}

For the numerical analysis of the problem it is more convenient to write Eqs.~(2.18)
in the form
\begin{mathletters}\label{3.1}
\begin{eqnarray}
\alpha_0&=&\sqrt{\frac{1+\sigma_1}{1+\sigma_0}}+c\alpha_0\sqrt{\alpha_0\alpha_1},\\
\alpha_1&=&\sqrt{\frac{1+\sigma_0}{1+\sigma_1}},\\
\sigma_0&=&-\frac{\alpha_1\sigma_1}{\alpha_0}+2c(1+\sigma_0)\sqrt{\alpha_0\alpha_1},\\
\sigma_1&=&\frac{\alpha_0^2(1+\sigma_0)}{\sqrt{\alpha_0\alpha_1}}b,
\end{eqnarray}
\end{mathletters}
where
\begin{equation}
c=\frac{(D-2)}{2\pi}\frac{\beta (s)}{\mu^2s}; 
\qquad \beta (s)=\frac{\partial\eta (s)}{\partial s},
\end{equation}
\begin{equation}
b=\frac{(D-2)}{2\pi}\frac{\eta (s)}{\mu^2s^2}.
\end{equation}
Combining Eqs.~(3.1a) and (3.1b) we get
\begin{equation}
\alpha_0\alpha_1=1+c\alpha_0\alpha_1\sqrt{\alpha_0\alpha_1},
\end{equation}
or
\begin{equation}
cx^3-x^2+1=0,
\end{equation}
where
\begin{equation}
x=\sqrt{\alpha_0\alpha_1}.
\end{equation}
We can now solve Eqs.~(3.1) in terms of $x$
\begin{mathletters}\label{3.7}
\begin{eqnarray}
\alpha_0&=&\sqrt{\frac{1+(b-2c)x}{(1-(b+c)x)(1-cx)}},\\
\alpha_1&=&\sqrt{\frac{1-(b+c)x}{(1+(b-2c)x)(1-cx)}},\\
\sigma_0&=&-\frac{(b-2c)x}{1+(b-2c)x},\\
\sigma_1&=&\frac{bx}{1-(b+c)x}.
\end{eqnarray}
\end{mathletters}
Thus at the end everything depends now on $s$ and $\mu$. From Eq.~(2.13) we can recover
the $\rho$-dependence. In figures $1$ to $3$ we have the behavior of $\eta$,
$\beta$, $c$ and $b$ as functions of $s$ for various values of the
mass parameter $\mu$. We see in Fig.1 that $\eta(s)$ presents a maximum for 
$s = s_0 \approx 0.27$, at this point $\beta = \frac{\partial\eta (s)}{\partial s}$ vanishes 
and the problem 
can be solved exactly. This point corresponds to the maximum value of the Casimir 
energy Eq.~(2.15) for a given length $\rho$. In this point $c(s_0) = 0$ and from 
Eq.~(3.5) $x = 1$. Thus the solution to the gap equations is
\begin{mathletters}\label{3.8}
\begin{eqnarray}
\alpha_0&=&\sqrt{1-2\lambda},\\
\alpha_1&=&\frac{1}{\sqrt{1-2\lambda}},\\
\sigma_0&=&\frac{\lambda}{1-2\lambda},\\
\sigma_1&=&-\lambda.
\end{eqnarray}
\end{mathletters}
From Eqs.~(2.13), (3.8a) and (3.8c) we see that
\begin{equation}
\lambda=1-\frac{\mu^2s_0^2}{\rho^2},
\end{equation}
comparing with Eq.~(2.15) we find
\begin{equation}
\rho=\mu s_0\sqrt{1-\frac{(D-2)}{2\pi}\frac{\eta (s_0)}{\mu^2s_0^2}}.
\end{equation}
The potential can then be written as a function of $\mu$ as follows
\begin{equation}
 \overline{V}(\mu)=\mu s_0\sqrt{1+\frac{(D-2)}{2\pi}\frac{\eta (s_0)}{\mu^2s_0^2}}.
\end{equation}
Since $\eta(s_0)<0$ Eq.~(3.11) implies that there is a minimum value of $\mu$
for which the potential exists at $s = s_0$, denoting this value by $\mu_{min}$
we see that it is given by
\begin{equation}
\mu_{min}=\sqrt{\frac{(D-2)}{2\pi}\frac{|{\eta (s_0)}|}{s_0^2}}.
\end{equation}
The general solution of the problem is given by Eqs.~(3.5) and (3.7). Fig.4
shows the behaviour of $\alpha_0$, $\alpha_1$, $\sigma_0$ and $\sigma_1$ as
functions of $\rho$ for various values of the mass parameter $\mu$. We see that
for large $\rho$, $\alpha_0$ and $\alpha_1$ tend to one while $\sigma_0$ and
$\sigma_1$ approach zero. Thus from Eq.~(2.13) we see that for finite $\mu$
large-$\rho$ is equivalent to large-$s$, this will be of interest when 
discussing L\"{u}scher term in the following section.
In Fig.5 we show the behaviour of the potential $\overline{V}(\rho)$ for several values 
of the mass $\mu$. We see that for big and small values of $\mu$ the curves
come close together in agreement with Eq.~(2.12) approaching the Nambu-Goto 
result for $\mu = 0, \infty$. We also see that the small bump in 
Fig.2 of [17] for $\mu \approx 0.3$ is not present. This being probably a numerical 
artefact. We next show in Fig.6 the so called deconfinement radious $\rho_{dec}$
as a function of $\mu$. This is the value of $\rho$ for which the potential 
vanishes and probably signals the presence of the tachyon in string models.
Comparing with Fig.3 of [17] we see that the behaviour is very similar avoiding,
however, the numerical trick of [17] at $\mu \approx 0.1$.

\section{Discussion and Conclusions}

We have obtained exact results to the problem of quark mass corrections to the
string potential for the Nambu-Goto model in the case where the masses attached
to the ends of the string are equal. These results are similar to those presented 
by Lambiase and Nesterenko [17] obtained under some symplifying assumptions.
There is, however, a subtle point concerning the L\"{u}scher term which we would like 
to discuss. For a string with fixed ends L\"{u}scher term has a contribution to the 
potential of the form
\begin{equation}
\overline{V}_L(\rho)=-\frac{(D-2)\pi}{24\rho}.
\end{equation}
The importance of this term is that it is universal i.e., independent of the 
details of a whole class of models, in particular, independent of the parameters 
of the model under consideration. In the one-loop approximation to the problem 
discussed above the potential becomes
\begin{equation}
\overline{V}(\rho)=\rho + (D-2)E_c,
\end{equation}
where $E_c$ is given by
\begin{equation}
E_c=\frac{\eta (s)}{2\pi\rho},
\end{equation}
and $E_c$ depends on the mass $\mu$ through $s$ (see Eq.~(2.13)) thus apparently
giving L\"{u}scher term a mass dependence. It is important to notice, however, that
this Coulomb-like term arises as a long distance (large-$\rho$) effect. Thus strictly
speaking corrections to L\"{u}scher term, if any, should be obtained after expanding
Eq.~(4.2) for large $\rho$. From our numerical results we can see that large-$\rho$
is equivalent to large-$s$ for a given finite value of $\mu$. Thus for large 
$\rho$, $\alpha_0$ and $\alpha_1$ are essentially one and from Eqs.~(2.20) and (2.21)
the potential becomes
\begin{equation}
\overline{V}(\rho)\approx\rho\sqrt{1-2\lambda}\approx\rho (1-\lambda+...).
\end{equation}
For large-s we can approximate the integral involved in the definition of $\lambda$
Eq.~(2.17) with the result
\begin{equation}
\lambda\approx\frac{(D-2)\pi}{24\rho^2}-\frac{(D-2)\pi}{12\rho^3}\mu, 
\qquad s \rightarrow \infty.
\end{equation}
Thus the potential becomes
\begin{equation}
\overline{V}(\rho)=\rho - \frac{(D-2)\pi}{24\rho}+\frac{(D-2)\pi}{12\rho^2}\mu+...
\end{equation}
leaving L\"{u}scher term universal.

In conclusion the study of the interquark potential for string models with 
masses attached to its ends is of undoubted interest by itself as a 
mathematical problem and certainly for the possible physical applications 
to the low energy regime of QCD.  Here we have presented exact solutions 
to the gap equations and the interquark potential has been obtained for 
several values of $\mu$. We see that having finite point masses
at the ends of the string has considerable effects on the potential. Also the
deconfinement radious become a function of $\mu$ and its value could be fixed
phenomenologically. We also discuss the universality of L\"{u}scher term and argue
that it remains universal if we understand it strictly as a long distance
effect with mass corrections coming up at higher orders in $\rho^{-1}$. Finally the
tachyon problem of string theories remains unresolved although recently [18] there 
has been some discussion on how one can possibly avoid it.
\newpage
\noindent
{\bf Figure Captions}
 
\bigskip
\noindent
Fig.1\\
We show  $\eta (s)$ as a function of log(s) Eq.~(2.16) which essentially defines the 
Casimir energy Eq.~(2.12). For $\mu = 0, \infty$ the quantity s given by 
Eq.~(2.13) takes the values $\infty, 0$ respectively and the Casimir energy becomes
$E_c = -\frac{\pi}{24R}$. This value coincide with the one obtained for a string with 
fixed ends boundary conditions. We see that $\eta (s)$ has a maximum at 
$s = s_0 \approx 0.27$. At this point we can obtain an exact analytical solution 
given by Eqs.~(3.8).

\bigskip
\noindent
Fig.2\\
The quantity $\beta=\frac{\partial\eta (s)}{\partial s}$ is shown as a function of s. 
The point $s = s_0 \approx 0.27$ where $\beta (s_0) = 0$ corresponds to the maxima
of $\eta(s)$. From Eq.~(3.2) we see that $c(s_0)=0$, Eq.~(3.5) implies $x=1$ and 
a particular solution follows (see Eqs.~(3.8)).

\bigskip
\noindent
Fig.3\\
We show the quantities denoted by $c(s)$ (dashed line) and $b(s)$ (solid line)
as functions of s for $\mu=10^{-3}$ (Fig.3a) and $10^3$ (Fig.3b). These quantities 
are defined by Eq.~(3.2) and (3.3) respectively. In Fig.3a $c(s)$ eventually
reaches a minimum value and then goes up passing through zero at $s = s_0$.
The curve for $b(s)$ is always negative as follows from Eq.~(3.3).

\bigskip
\noindent
Fig.4\\
The solutions to the gap equations Eqs.~(3.7) for the Lagrange 
multipliers $\alpha_0$, $\alpha_1$ and metric components $\sigma_0$, $\sigma_1$
are shown as functions of $\rho = M_0R$ for $\mu=100,1$, and $0.1$ 
(solid, dashed and dash-dotted lines, respectively). In Fig.4a the curves for 
$\alpha_0$ are the lower ones and for $\alpha_1$ the upper ones, while those for 
$\sigma_0$ (above) and $\sigma_1$ (below) appear in Fig.4b. 
The minimum value $\alpha_0$ can reach is zero as follows from Eq.~(2.21).

\bigskip
\noindent
Fig.5\\
The dimensionless interquark potential $\overline{V}(\rho)$ Eq.~(2.20) is here shown as a
function of $\rho = M_0R$ for $\mu=0.3, 10, 100, 0$ or $\infty$ (dotted, dashed,
solid and dash-dotted lines, respectively). When 
$\mu = 0, \infty$ (dash-dotted line) corresponding to free and fixed ends strings respectively the potential 
becomes the well known Nambu-Goto potential. As the mass $\mu$ varies between 
zero and infinity $\overline{V}(\rho)$ essentially keeps its shape but reaching a vanishing 
value at different deconfinement radious $\rho_{dec}$ (see Fig.6). In all the
cases the potential becomes linear for large values of $\rho$.

\bigskip
\noindent
Fig.6\\
The so called deconfinement radious $\rho_{dec}$ which is defined as the value 
of $\rho$ for which $\overline{V}(\rho = \rho_{dec}) = 0$ is here shown as a function of 
the mass parameter $\mu = \frac{m}{M_0}$. 
For $\mu = 0, \infty$ $\rho_{dec}(\mu) = \sqrt{\frac{(D-2)\pi}{12}}|_{D=4}
\approx 0.72$ as in 
the Nambu-Goto 
case. For finite $\mu$-values $\rho_{dec}$ lies in the
interval $0.31\leq \rho_{dec} \leq \sqrt{\frac{(D-2)\pi}{12}}|_{D=4}\approx 0.72$

\newpage
{\bf References }

\medskip
\begin{itemize}
\item[1)] Y. Nambu, in {\it Symmetries and Quark Models}, edited by R. Chand 
(Gordon and Breach, New York, 1970); T. Goto, Prog. Theor. Phys. {\bf 46}, 
1560(1971).

\item[2)] T. Eguchi, Phys. Rev. Lett. {\bf 44}, 126(1980).

\item[3)] A. Schild, Phys. Rev. D{\bf 16}, 1722(1977).

\item[4)] M. L\"{u}scher, K. Symanzik, and P. Weisz, Nucl. Phys. {\bf B173}, 365(1980),

\item[5)] O. Alvarez, Phys. Rev. D{\bf 24}, 440(1981).

\item[6)] J.F. Arvis, Phys. Lett {\bf 127B}, 106 (1983).
 
\item[7)] R.D. Pisarski and O. Alvarez, Phys. Rev. D{\bf 26}, 3735(1982); 
A. Antill\'on and G. Germ\'an, Phys. Rev. D{\bf 47}, 4567(1993).

\item[8)] M. Flensburg and C. Peterson, Nucl. Phys. {\bf B283}, 141(1987);
F. Karsch and E. Laermann, Rep. Prog. Phys. {\bf 56}, 1347(1993).

\item[9)] J. Polchinski, Phys. Rev. Lett. {\bf 68}, 1267(1992); Phys. Rev.
D{\bf 46}, 3667(1992).

\item[10)] G. Germ\'an, H. Kleinert, and M. Lynker, Phys. Rev. D{\bf 46}, 
1699 (1992); G. Germ\'an, M. Lynker, and A. Mac\'{\i}as, Phys. Rev. D{\bf 46},
3640(1992); G. Germ\'an and M. Lynker, Phys. Rev. D{\bf 46}, 5678(1992);
A. Antill\'on and G. Germ\'an, Phys. Rev. D{\bf 49}, 1966(1994).

\item[11)] M. Natsuume, Phys. Rev. D{\bf 48}, 835(1993).

\item[12)]  A.M. Polyakov, Nucl. Phys. {\bf B268}, 406(1986); H. Kleinert, 
Phys. Lett. {\bf 174B} 335(1986); for a review see A. M. Polyakov, 
{\it Gauge Fields and Strings} (Harwood Academic Publishers, Chur, 1987). 

\item[13)] G. Germ\'an, Mod. Phys. Lett. A{\bf 6}, 1815(1991).

\item[14)] M. Awada, Phys. Lett. {\bf 351B}, 468(1995).

\item[15)]  A.M. Polyakov, PUPT-1632, hep-th/9607049; M.C. Diamantini, 
F. Quevedo, and C.A. Trugenberger, Phys. Lett. {\bf 396B}, 115(1997).

\item[16)] V.V. Nesterenko, Z. Phys. C{\bf 51}, 643(1991); V.V. Nesterenko and 
N.R. Shvetz, Z. Phys. C{\bf 55}, 265(1992); B.M. Barbashov and V.V. Nesterenko, 
{\it Introduction to the Relativistic String Theory} (Worl Scientific, Singapore, 1990).

\item[17)] G. Lambiase and V.V. Nesterenko, Phys. Rev. D{\bf 54}, 6387(1996).
 
\item[18)] H. Kleinert, G. Lambiase , and V.V. Nesterenko, Phys. Lett. 
{\bf 384B}, 213(1996).

\end{itemize}

\end{document}